\documentclass[a4paper,fleqn]{cas-sc}

\usepackage[numbers]{natbib}
\usepackage{mathtools}
\usepackage{algorithm}
\usepackage{algorithmic}
\renewcommand{\algorithmicrequire}{\textbf{Input:}}

\def\tsc#1{\csdef{#1}{\textsc{\lowercase{#1}}\xspace}}
\tsc{WGM}
\tsc{QE}
\tsc{EP}
\tsc{PMS}
\tsc{BEC}
\tsc{DE}
\begin{document}
\let\WriteBookmarks\relax
\def\floatpagepagefraction{1}
\def\textpagefraction{.001}
\shorttitle{Fast Classical Simulation by Simultaneous Diagonalization}
\shortauthors{Y.~Kawase and K.~Fujii}
%\begin{frontmatter}

% title 変更を考える
\title [mode = title]{
Fast Classical Simulation of Hamiltonian Dynamics by Simultaneous Diagonalization 
Using Clifford Transformation with Parallel Computation
}
%\tnotemark[1,2]

%\tnotetext[1]{This document is the results of the research
%   project funded by the National Science Foundation.}

%\tnotetext[2]{The second title footnote which is a longer text matter
%   to fill through the whole text width and overflow into
%   another line in the footnotes area of the first page.}

\author[1]{Yoshiaki Kawase}[type=editor]
\fnmark[1]
\ead{yoshiaki.kawase@qc.ee.es.osaka-u.ac.jp}

% \credit{Conceptualization of this study, Methodology, Software}

\address[1]{
Graduate School of Engineering Science, Osaka University,
1-3 Machikaneyama, Toyonaka, Osaka 560-831, Japan}

\author[1,2,3]{Keisuke Fujii}[type=editor]
\cormark[1]
\fnmark[2]
\ead{fujii@qc.ee.es.osaka-u.ac.jp}

%\credit{Data curation, Writing - Original draft preparation}

\address[2]{
Center for Quantum Information and Quantum Biology, Osaka University, Japan.
}
\address[3]{RIKEN Center for Quantum Computing, Wako Saitama 351-0198, Japan.}

%\author[1,3]{Sample.}
%\cormark[2]
%\fnmark[1,3]
%\ead{rishi@stmdocs.in}
%\ead[URL]{www.stmdocs.in}

\cortext[cor1]{Corresponding author}
%\cortext[cor2]{Principal corresponding author}
%\fntext[fn1]{This is the first author footnote. but is common to third author as well.}
%\fntext[fn2]{Another author footnote, this is a very long footnote and
%  it should be a really long footnote. But this footnote is not yet
%  sufficiently long enough to make two lines of footnote text.}

%\nonumnote{This note has no numbers. In this work we demonstrate $a_b$
%  the formation Y\_1 of a new type of polariton on the interface
%  between a cuprous oxide slab and a polystyrene micro-sphere placed
%  on the slab.
%  }

\begin{abstract}
Simulating quantum many-body dynamics is important both for fundamental understanding of physics and practical applications for quantum information processing.
Therefore, classical simulation methods have been developed so far.
Specifically, the Trotter-Suzuki decomposition can analyze a highly complex quantum dynamics,
if the number of qubits is sufficiently small so that main memory can store the state vector.
However,
simulation of quantum dynamics via Trotter-Suzuki decomposition requires 
huge number of steps,
each of which accesses the state vector, 
and hence the simulation time becomes impractically long.
To settle this issue,
we propose a technique to accelerate simulation of quantum dynamics via simultaneous diagonalization of mutually commuting Pauli groups,
which is also attracting a lot of attention to reduce the measurement overheads in quantum algorithms.
We group the Hamiltonian into mutually commputing Pauli strings, and each of them are diagonalized in the computational basis via a Clifford transformation.
Since diagonal operators are applied on the state vector simultaneously with minimum memory access,
this method successfully use performance of highly parallel processors such as Graphics Processing Units (GPU). 
Compared to an implementation using one of the fastest simulators of quantum computers, the numerical experiments have shown that our method provides a few tens of times acceleration.
\end{abstract}

%\begin{graphicalabstract}
%\includegraphics{imgs/ising_cpu_gpu.png}
%\end{graphicalabstract}

%\begin{highlights}
%\item Research highlights item 1
%\item Research highlights item 2
%\item Research highlights item 3
%\end{highlights}


\maketitle

\section{Introduction}

Quantum many-body dynamics has been studied from both a practical perspective, such as its application to quantum information processing, and from a fundamental physical interest
such as quantum chaos \cite{joshi2022probing}, quantum information scrambling \cite{swingle2016measuring,landsman2019verified}, and discrete-time crystal \cite{zhang2017observation}. 
While controlled quantum dynamics has been experimentally realized in various physical systems, classical simulation of quantum dynamics is still necessary to investigate detailed properties of quantum many-body dynamics and to validate experimental data.
As pointed out by R. Fenyman, classcial simulation of quantum many-body dynamics is in general hard for classical computers,
which is the reason why sophisticated classical simulation methods, such as
exact diagonalization, full vector simulaiton, tensornetwork approximation,
have been developped.
An exact diagonalization incurs high memory cost and execution time 
because the dimensions of the quantum system exponentially increase with the number of qubits. 
Moreover, for a time dependent Hamiltonian, diagonalization does not allow us to calculate the time evolution.
A simulation based on tensor network
can efficiently simulate a low-depth quantum circuit or quantum computation with relatively small entanglement growth \cite{villalonga2019flexible}. 
A quantum circuit to simulate highly complex quantum many-body dynamics with a large depth may not satisfy such conditions. 
Alternatively, we can use the full vector simulation,
where 
quantum dynamics is simulated 
by updating a quantum state 
with infinitesimal time evolution operators
obtained by Trotter-Suzuki decomposition \cite{suzuki1991general}. 
Once we have enough memory to represent a quantum state, 
we can simulate the time evolution by updating the quantum state using a sparse matrix,
without any diagonalization. 
This approach is most advantageous, 
as long as the system size stays within the limitation by memory, 
since the overhead scales linearly in time.

However, in practice, the Trotter-Suzuki decomposition requires at least hundreds or thousands steps to achieve sufficient accuracy for simulation,
which results in long simulation time and makes this approach impractical for a large system even within the memory size.
To settle this,
parallel computing architectures can be employed to accelerate 
simulation. 
One of the most ubiquitous parallel computing architectures is Graphics Processing Units (GPU), which has thousands of threads.
In Ref.~\cite{dente2013gpu}, the simulation time of the time evolution of quantum spin dynamics using the Trotter-Suzuki decomposition is $30$ times faster by GPU than CPU. 
However, this approach only applicable a simple Hamiltonian and cannot be 
applied for a general Hamiltonian including an arbitrary Pauli strings.
On the other hand, GPU-based simulation has proven its usefulness in simulating discretized
quantum time evolution for quantum circuits in quantum computing \cite{suzuki2020qulacs,Qiskit},
while it is not optimized for simulating continuous quantum time evolution.

In this paper, 
by focusing on Hamiltonian dynamics of quantum many-body systems, we propose a general method to accelerate simulations and show that 
real-time dynamics of quantum many-body systems can be simulated in realistic time even for $30$ qubit-scale ($10^9$ dimensional state vector) quantum dynamics where the size of the GPU memory permits.
To this end, 
the bottleneck is the huge number of steps originated by 
discretization via the Trotter-Suzuki decomposition for continuous time evolution as mentioned above.
To relax this bottleneck, 
we diagonalize a part of Hamiltonian so that 
multiple updates on a quantum state are done at once with minimal memory access.
Since we do not diagonalize whole Hamiltonian, the diagonalization 
can be done efficiently and systematically. 

More specifically,
in our method, Hamiltonian is partitioned into groups of mutually commuting Pauli terms, and each group is transformed by a unitary transformation consisting of Clifford gates to a diagonal matrix. 
Interestingly, exactly the same task is being investigated as a way to improve the efficiency of quantum algorithms.
More precisely, 
the variational quantum eigensolver \cite{peruzzo2014variational},
which is one of the most promising quantum algorithms that 
works on near-term quantum computing devices,
estimates an expectation value of Hamiltonian using a quantum computer.
There, simultaneous diagonalization of mutually commuting Pauli terms has been employed to reduce the measurement complexity \cite{gokhale2019minimizing}. 
Furthermore, a similar technique is also applied 
to reduce the CNOT gates in simulation of Hamiltonian dynamics on actual quantum devices \cite{berg2020circuit}. 
All these are intended to reducing quantum computational overhead including quantum gates and measurements.

In this work, we apply and optimize Hamiltonian partitioning technique for the purpose of accelerating classical simulation of Hamiltonian dynamics 
by simultaneously diagonalizing the mutually commuting Pauli terms.
 A matrix exponential of the diagonal matrix corresponding to mutually commuting Pauli terms is easily calculated and applied to a quantum state at once with minimum memory access.
 Note that without this simultaneous diagonalization,
the time evolution operator must be applied for each term in the Hamiltonian with accessing the state vector each time.
This simultaneous diagonalization allows us to reduce the number of steps required to simulate the Hamiltonian dynamics and overhead for accessing the exponentially large dimensional state vector.
Thereby, the simulation time of the Hamiltonian dynamics
is much reduced in total even though an additional computational cost is required to perform the unitary transformation for diagonalization.

To verify our algorithm and analyze the performance, 
we perform numerical experiments. 
In our numerical experiments, we simulate the time evolution under the Hamiltonian of transverse field Ising model and SYK Majorana fermion model \cite{sachdev1993gapless,kitaev2015simple}.
The former, the transverse field Ising model is a prototypical model of quantum spin systems and is also widely employed for quantum information processing such as quantum annealing \cite{kadowaki1998quantum} and quantum approximated optimization algorithm \cite{farhi2014quantum}.
The latter is another prototypical model for a highly complex quantum many-body system, which exhibits quantum informational scrambling \cite{iyoda2018scrambling}. 
We implement the proposed method for these Hamiltonians
and compare the simulation time 
with straightforward Trotter-Suzuki decomposition simulated by fast and general purpose quantum computer simulator, Qulacs using both CPU and GPU \cite{suzuki2020qulacs}.
We confirm that the simulation is accelerated with our proposed method regardless of the architectures.
The former model is $10$ times faster, and the latter is two times faster by our proposed method even when the number of qubits is 30.
Furthermore, if the number of qubits is much smaller such as 15 qubits, 30 and 10 times faster, respectively.

The rest of this paper is organized as follows. 
In Sec.~\ref{sec:background}, we explain the basic elements of the classical simulation of quantum dynamics, and the binary tableau for simultaneous diagonalization.
Sec.~\ref{sec:simultaneous_diagonalization} describes our proposed method to simultaneously diagonalize mutually commuting terms and the classical simulation by simultaneous diagonalization.
To verify our proposed method, we perform numerical experiments in Sec.~\ref{sec:numerical_experiments}.
We conclude this work in Sec.~\ref{sec:conclusion}.

\section{Preliminary} \label{sec:background}
\subsection{Description of Quantum State and Quantum Gates} \label{subsec:basic_depcription_qc}
In a quantum computer, a qubit is the counterpart of a bit in a classical computer. 
A single-qubit quantum state is represented by a vector (state vector)
\begin{eqnarray} \label{eqn:state_vec_singlequbit}
|\psi \rangle = \alpha_0 | 0 \rangle + \alpha_1 | 1 \rangle \nonumber, 
\quad \text{where} \; |0\rangle=
\begin{pmatrix}
1 \\
0 
\end{pmatrix},
|1\rangle=
\begin{pmatrix}
0 \\
1 
\end{pmatrix},
\end{eqnarray}
where $\alpha_0$ and $\alpha_1$ are complex numbers, and these satisfy $|\alpha_0|^2+|\alpha_1|^2=1$.
%$|0 \rangle$ and $|1 \rangle$ represent row vectors. 
When we measure the qubit, we get an outcome $0$ with probability $|\alpha_0|^2$ and $1$ with probability $|\alpha_1|^2$.
The probability amplitudes of a quantum state with $n$ qubits are described by a complex vector of $2^n$ dimensions:
\begin{equation} \label{eqn:state_vec_nqubit}
|\psi \rangle = \alpha_{0 \ldots 00} | 0 \ldots 00 \rangle 
+ \alpha_{0 \ldots 01} | 0 \ldots 01 \rangle 
+ \cdots 
+ \alpha_{1 \ldots 11} | 1 \ldots 11 \rangle 
,
\end{equation}
where
\begin{equation}
|0 \ldots 00 \rangle=
\begin{pmatrix}
1 \\
0 \\
\vdots \\
0 
\end{pmatrix},
|0 \ldots 01\rangle=
\begin{pmatrix}
0 \\
1 \\
\vdots \\
0 
\end{pmatrix}
, \cdots,
|1 \ldots 11\rangle=
\begin{pmatrix}
0 \\
0 \\
\vdots \\
1 
\end{pmatrix}
\nonumber.
\end{equation}
An operation on qubits are represented by a unitary matrix. % updated by applying a quantum gate. 
A single-qubit gate is represented by a two-by-two unitary matrix. 
When we apply it on the $j$th target qubit, the quantum state is updated by the following equation 
for all $2^{n-1}$ combinations of $i_k \in \{0, 1\}, k=1,\ldots,j-1,j+1,\ldots,n$:
\begin{equation} \label{eqn:update_by_u1}
\left(
\begin{array}{cc}
\alpha_{i_n\ldots i_{j+1} 0 i_{j-1} \ldots i_1}^\prime \\
\alpha_{i_n\ldots i_{j+1} 1 i_{j-1} \ldots i_1}^\prime
\end{array}
\right)
= \mathit{U_1} \left (
\begin{array}{cc}
\alpha_{i_n\ldots i_{j+1} 0 i_{j-1} \ldots i_1} \\
\alpha_{i_n\ldots i_{j+1} 1 i_{j-1} \ldots i_1}
\end{array}
\right), 
\end{equation}
where the $\alpha_i^\prime$ denotes the updated amplitudes of $i$th basis state 
and ${\it U}_1$ a two-by-two unitary matrix. 
The examples of the matrices are Pauli matrices as follows:
\begin{equation} \label{eqn:pauli_matrix}
\sigma_0 = \mathit{I} \coloneqq \left(
\begin{array}{cc}
1 & 0 \\
0 & 1
\end{array}
\right),
\;
\sigma_1 = \mathit{X} \coloneqq \left(
\begin{array}{cc}
0 & 1 \\
1 & 0
\end{array}
\right),
\;
\sigma_2 = \mathit{Y} \coloneqq \left(
\begin{array}{cc}
0 & -i \\
i & 0
\end{array}
\right),
\;
\sigma_3 = \mathit{Z} \coloneqq \left(
\begin{array}{cc}
1 & 0 \\
0 & -1
\end{array}
\right). 
\end{equation}
Similarly, a two-qubit gate is defined by a four-by-four matrix. 
When we apply the gate on the $j$th and $k$th qubits as the target qubits, 
%updating the quantum state with this matrix is performed as follows.
the quantum state is updated by the following equation for all $2^{n-2}$ combinations of $i_l \in \{0, 1\},$ where $l$ satisfies $1\leq l \leq n$ and $l \neq j, k$:
\begin{eqnarray} \label{eqn:update_by_u2}
\left (
\begin{array}{cc}
\alpha_{i_n\ldots i_{j+1} 0 i_{j-1} \ldots i_{k+1} 0 i_{k-1} \ldots i_1}^\prime \\
\alpha_{i_n\ldots i_{j+1} 0 i_{j-1} \ldots i_{k+1} 1 i_{k-1} \ldots i_1}^\prime \\
\alpha_{i_n\ldots i_{j+1} 1 i_{j-1} \ldots i_{k+1} 0 i_{k-1} \ldots i_1}^\prime \\
\alpha_{i_n\ldots i_{j+1} 1 i_{j-1} \ldots i_{k+1} 1 i_{k-1} \ldots i_1}^\prime
\end{array}
\right)
= \mathit{U_2} \left (
\begin{array}{cc}
\alpha_{i_n\ldots i_{j+1} 0 i_{j-1} \ldots i_{k+1} 0 i_{k-1} \ldots i_1} \\
\alpha_{i_n\ldots i_{j+1} 0 i_{j-1} \ldots i_{k+1} 1 i_{k-1} \ldots i_1} \\
\alpha_{i_n\ldots i_{j+1} 1 i_{j-1} \ldots i_{k+1} 0 i_{k-1} \ldots i_1} \\
\alpha_{i_n\ldots i_{j+1} 1 i_{j-1} \ldots i_{k+1} 1 i_{k-1} \ldots i_1}
\end{array}
\right), 
\end{eqnarray}
where ${\it U}_2$ is a four-by-four matrix. 

A special class of unitary operators (or transformation) that 
maps a tensor product of Pauli matrices to 
another or the same tensor product of Pauli matrices is called Clifford operators (or transformations).
This kind of transformation includes the following quantum gates.
\begin{eqnarray} \label{eqn:clifford_gates}
\mathit{S} = \left(
\begin{array}{cc}
1 & 0 \\
0 & i
\end{array}
\right),
\quad
\mathit{H} = \frac{1}{\sqrt{2}}\left(
\begin{array}{cc}
1 & 1 \\
1 & -1
\end{array}
\right),
\quad
\text{CNOT} = \left(
\begin{array}{cccc}
1 & 0 & 0 & 0 \\
0 & 1 & 0 & 0 \\
0 & 0 & 0 & 1 \\
0 & 0 & 1 & 0
\end{array}
\right),
\quad
\text{CZ} = \left(
\begin{array}{cccc}
1 & 0 & 0 & 0 \\
0 & 1 & 0 & 0 \\
0 & 0 & 1 & 0 \\
0 & 0 & 0 & -1
\end{array}
\right). 
\end{eqnarray}
We show the example of transformation of the Pauli terms 
by the Clifford transformation in Table.~\ref{table:single_qubit_gate} and \ref{table:two_qubit_gate}.
Note that we denote a single-qubit gate ${\it V}$ on $t$th target qubit as ${\it V}_t$ and 
CNOT gate on $c$th control qubit and $t$th target qubit as CNOT$(c,t)$. 

The Clifford operations (or transformations)
play a very important role in quantum computing, 
since this class of operations can be efficiently simulatable on a classical computer by virtue of Gottesman-Knill theorem \cite{aaronson2004improved}.
In our method, 
we use the Clifford tansformation to partially diagonalize a Hamiltonian.

\begin{table}
	\caption{
	This table shows the transformation of Pauli matrices 
	by single-qubit gates in Clifford transformation. 
	We can see that the Pauli matrices turn into the Pauli matrices by H and S gates. 
	Note that we need to multiply $-1$ to transform the {\it X} or {\it Y} into {\it Y}.
	}
	\begin{tabular}{|c|c|c|} \hline
	   Pauli term & $U=H$              & $U=S$  \\ \hline
	   $X=$ & $U^\dagger Z U$    & $U^\dagger Y U$    \\ \hline
	   $Y=$ & $U^\dagger (-Y) U$ & $U^\dagger (-X) U$    \\ \hline
	   $Z=$ & $U^\dagger Z U$    & $U^\dagger Z U$    \\ \hline 
	\end{tabular}
	\label{table:single_qubit_gate}
\end{table}

\begin{table}
	\caption{
	This table shows the transformation of Pauli terms 
	by two-qubit gates in Clifford transformation. 
	We can see that the Pauli terms are transformed into Pauli terms by CNOT and CZ gates.
	Note that we need to multiply $-1$ to transform the ${\it X_1 Z_2}$ into ${\it Y_1 Y_2}$.
	}
	\begin{tabular}{|c|c|c||c|c|c|} \hline
	 Pauli terms & $U=\text{CNOT}(1,2)$ & $U=\text{CZ}(1,2)$ & 
	 Pauli terms & $U=\text{CNOT}(1,2)$ & $U=\text{CZ}(1,2)$ \\ \hline
	 $X_1 I_2 =$ & $U^\dagger X_1 X_2 U$ & $U^\dagger X_1 Z_2 U$ &
	 $Y_1 I_2 =$ & $U^\dagger Y_1 X_2 U$ & $U^\dagger Y_1 Z_2 U$ \\ \hline
	 $I_1 X_2 =$ & $U^\dagger I_1 X_2 U$ & $U^\dagger Z_1 X_2 U$ &
	 $I_1 Y_2 =$ & $U^\dagger Z_1 Y_2 U$ & $U^\dagger Z_1 Y_2 U$ \\ \hline
	 $X_1 X_2 =$ & $U^\dagger X_1 I_2 U$ & $U^\dagger Y_1 Y_2 U$ &
	 $Y_1 Y_2 =$ & $U^\dagger (-X_1 Z_2) U$ & $U^\dagger X_1 X_2 U$ \\ \hline 
	\end{tabular}
	\begin{tabular}{|c|c|c|} \hline
	 Pauli terms & $U=\text{CNOT}(1,2)$ & $U=\text{CZ}(1,2)$ \\ \hline 
	 $Z_1 I_2 =$ & $U^\dagger Z_1 I_2 U$ & $U^\dagger Z_1 I_2 U$ \\ \hline
	 $I_1 Z_2 =$ & $U^\dagger Z_1 Z_2 U$ & $U^\dagger I_1 Z_2 U$ \\ \hline
	 $Z_1 Z_2 =$ & $U^\dagger I_1 Z_2 U$ & $U^\dagger Z_1 Z_2 U$ \\ \hline
	\end{tabular}
	\label{table:two_qubit_gate}
\end{table}

\subsection{Classical Simulation of Quantum Dynamics}\label{subsec:classical_simulation_qd}
The time evolution of a quantum state within a time-independent Hamiltonian $H$ is described by the Schr\"{o}dinger equation, and the solution is written by $|\psi(t) \rangle = e^{-iHt} |\psi(0) \rangle$, where $|\psi(0) \rangle$ is the initial state. 
Since the dimensions of the Hamiltonian exponentially increases with the size of a quantum system, the simulation is intractable in general. 
However, the Hamiltonian usually can be written as a sum over a polynomial number of  local interactions with respect to the system size. 
Let $H_j$ be a local Pauli operators,
whose support is a finite number of qubits, with a certain coefficient.
We rewrite the Hamiltonian by
\begin{equation*} 
H=\sum_{j=1}^m c_j H_j 
\qquad
\text{where} \quad c_j \in \mathbb{R}, \, H_j\in\{I,X,Y,Z\}^{\otimes n}, 
\end{equation*}
and, $m$ denotes the number of terms of the Hamiltonian. 
Under the Hamiltonian, we approximate the time evolution by Trotter-Suzuki decomposition \cite{suzuki1991general}. 
More precisely, the quantum state at time $t$ can be written as 
\begin{eqnarray} \label{eqn:trotter_decomp}
|\psi(t) \rangle = e^{-iHt} |\psi(0) \rangle
 = \lim_{n_{TS} \to \infty} \left( \Pi_{j=1}^m e^{-i c_j H_j (t/n_{TS})} \right)^{n_{TS}} | \psi(0) \rangle
 \simeq \left( \Pi_{j=1}^m e^{-i c_j H_j \Delta t} \right)^{n_{TS}} | \psi(0) \rangle ,
\end{eqnarray}
where the $n_{TS}$ denotes the number of Trotter steps and the $\Delta t=t/n_{TS}$ is the time step of the Trotter-Suzuki decomposition.
The simulation of infinitesimal time evolution operator $e^{-i c_j H_j \Delta t}$ by multiplying the state vector by the unitary matrix, such as Eq.~\eqref{eqn:update_by_u1} and \eqref{eqn:update_by_u2},
where we use the fastest quantum circuit simulator ``qulacs'' \cite{suzuki2020qulacs}.

\subsection{Partitioning Hamiltonian Terms} \label{subsec:partitioning_hamiltonian}

In the simulation of Hamiltonian dynamics with Trotter-Suzuki decomposition as Eq.~\eqref{eqn:trotter_decomp}, we multiply the state vector by a unitary matrix for each Trotter step. 
In the multiplication, the state vector must be read multiple times from the main memory 
if the size of the state vector is larger than that of the cache. 
Since the simulation time includes the time for the multiplication and memory access, 
we accelerate the simulation by reducing the number of memory access.
To this end, once we read the amplitudes, 
we simultaneously multiply the exponentiation of multiple Hamiltonian terms. 
More precisely, we partition the whole Hamiltonian terms into mutually commuting terms, and 
simultaneously time-evolving each group of the partitioned Hamiltonian with diagonalization. 

Interestingly the same task is also done in 
a different context to reduce the number of measurements in variational quantum eigensolver (VQE) \cite{gokhale2019minimizing}.
More precisely, 
the expectation value of the whole Hamiltonian terms in the group can be calculated 
from that of the part of the terms by diagonalizing the group of mutually commuting terms. 
Similarly, 
we diagonalize groups of mutually commuting terms and simultaneously time-evolve the terms in each group
to accelerate the classical simulation.
Since the simulation time depends on the number of groups, 
it is important to decrease the number of groups to accelerate the simulation. 
However, finding the partition of given Hamiltonian with the minimum number of groups is an NP-hard problem. 
To avoid this, we obtain the approximation of the partition by greedy method. 

\subsection{Tableau Representation of Pauli Operators and Clifford Operations}
\label{subsec:binary_tableau}
In order to perform a faster classical simulation, we have to find the Clifford transformation 
that diagonalizes the mutually commuting Pauli terms efficiently. 
To this end, we make a tableau consisting of the binary representation of Pauli terms, and search the Clifford transformation by applying it to the tableau systematically.
Once we find the Clifford transformation, we can also know the diagonalized terms. 
Here we explain the binary representation of Pauli terms and how the Clifford transformation acts on them.

First we describe the binary representation of Pauli terms in the Hamiltonian. 
Suppose the terms in the $j$th group of mutually commuting terms are written by 
\begin{equation} \label{eqn:pauli_term}
    P_j = \sum_{k=1}^{m_j} c_{j,k} P_{j,k}, \qquad \text{where} \; c_{j,k} \in \mathbb{R}, P_{j,k} \in \{I,X,Y,Z\}^{\otimes n},
\end{equation}
$m_j$ is the number of terms in $j$th group. 
Since $Y=i {\it X Z}$ holds, 
each Pauli term $P_{j,k}$ is also written by 
\begin{equation*}
    P_{j,k}=i^{s_k} \otimes_{l=1}^n {\it X}^{x_{k,l}} {\it Z}^{z_{k,l}}, 
    \qquad \text{where} \; s_k \in \{0,1,2,3\}, x_{k,l}\in\{0,1\}, z_{k,l}\in\{0,1\}.
\end{equation*}
We represent the terms in $P_{j,k}$ by $x_{k,l}$ and $z_{k,l}$ 
as the length $2n$ of bit string $x_{k,1} \ldots x_{k,n} z_{k,1} \ldots z_{k,n}$. 
For example, ${\it X_1 Y_2 Z_3}$ is represented by $110011$. 
By arranging the bit strings, 
we can make a tableau consisting of $m_j$ rows and $2n$ columns as
\begin{equation*}
\left[
\begin{array}{ccc|ccc}
x_{1,1} & \cdots & x_{1,n} & z_{1,1} & \cdots & z_{1,n} \\
\vdots & \ddots & \vdots & \vdots & \ddots & \vdots \\
x_{m_j,1} & \cdots & x_{m_j,n} & z_{m_j,1} & \cdots & z_{m_j,n} \\
\end{array}
\right].
\end{equation*}
%where $x_{k,l}\in \{ 0,1\}$ and $z_{k,l} \in \{0,1\}$. 
Note that we say a set of binary strings are linearly independent 
if each of the binary strings in the set cannot be represented by exclusive-OR between the other binary strings. 
The coefficient $i^{s_k}$ is caused by Pauli {\it Y}. 
From the Table \ref{table:single_qubit_gate} and \ref{table:two_qubit_gate}, 
the coefficient whose original Pauli term has Pauli {\it Y} turns into $+1$ or $-1$ by Clifford transformation.
So, in the classical simulation, 
we only need to track the sign. 
We add the column in $2n+1$th column of the tableau, 
and initialize the elements $r_k$ with $r_k=0$.
In this way, we obtain the tableau 
\begin{equation} \label{eqn:binary_tableau}
\left[
\begin{array}{ccc|ccc|c}
x_{1,1} & \cdots & x_{1,n} & z_{1,1} & \cdots & z_{1,n} & r_1 \\
\vdots & \ddots & \vdots & \vdots & \ddots & \vdots & \vdots \\
x_{m_j,1} & \cdots & x_{m_j,n} & z_{m_j,1} & \cdots & z_{m_j,n} & r_{m_j} \\
\end{array}
\right],
\end{equation}
where
% $x_{k,l}\in \{ 0,1\}$, $z_{k,l} \in \{0,1\}$ and
$r_k \in \{ 0,1\}$.
We call the first $n$ columns of the tableau as {\it X} block 
and the following $n$ columns of the tableau as {\it Z} block. 

\begin{figure}
	\centering
		\includegraphics[width=0.95\linewidth]{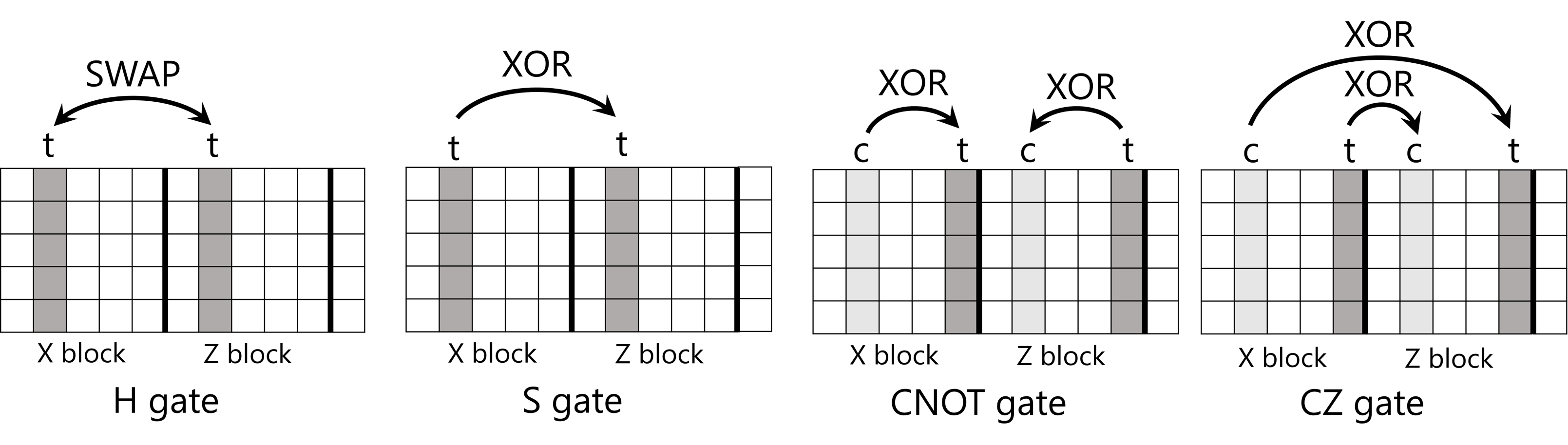}
	\caption{
	This figure shows the Clifford operations on the binary tableau of {\it X} and {\it Z} blocks.
	In the figure, each $t$ and $c$ represents target and control qubits. 
	We represent the $t$th column with dark gray and $c$th column with ligth gray. 
	}
	\label{fig:clifford_ope}
\end{figure}

Next, we explain how to update the tableau 
under the action of each elementary Clifford transformation~\cite{berg2020circuit,aaronson2004improved}. 
We summarize them in Table~\ref{gate_table}, 
and depict the image of the transformation in Fig.~\ref{fig:clifford_ope}. 
The binary tableau is updated by each Clifford gate according to the following procedure:
\begin{enumerate}
    \item {\it H gate}\\
    If we apply {\it H} gate on $t$th qubit, we update $r_k$ by $r_k= r_k \oplus x_{k,t} z_{k,t}$.
    Then we update the {\it X} block and {\it Z} block by swapping $x_{k,t}$ and $z_{k,t}$ for all $k=1,\ldots,m_j$.\\

    \item {\it S gate}\\
    If we apply {\it S} gate on $t$th qubit, 
    we update $r_k$ by $r_k= r_k \oplus x_{k,t} z_{k,t}$.
    Then we update the {\it Z} block by $z_{k,t} = z_{k,t} \oplus x_{k,t} $ for all $k=1,\ldots,m_j$.\\

    \item {\it CNOT gate}\\
    If we apply CNOT gate on $c$th control qubit and $t$th target qubit, 
    we update $r_k$ by $ r_k = r_k \oplus x_{k,c} z_{k,t} (x_{k,t} \oplus z_{k,c} \oplus 1) $.
    Then we update the {\it X} block and {\it Z} block by $x_{k,t}=x_{k,t} \oplus x_{k,c}$ and $z_{k,t} = z_{k,t} \oplus z_{k,c} $ for all $k=1,\ldots,m_j$.\\

    \item {\it CZ gate}\\
    If we apply CZ gate on $c$th control qubit and $t$th target qubit, 
    we update $r_k$ by $r_k = r_k \oplus x_{k,c} x_{k,t} (z_{k,t} \oplus z_{k,c} ) $.
    Then we update the {\it Z} block by $z_{k,t}=z_{k,t} \oplus x_{k,c}$ and $z_{k,c} = z_{k,c} \oplus x_{k,t} $ for all $k=1,\ldots,m_j$.
    
    % \item {\it updating $r_i$}\\
    %\item {\it updating X and Z blocks}\\
\end{enumerate}
In the above procedure, we denote the exclusive-OR as $\oplus$.
These operations can be derived from the Clifford operations on Pauli terms 
in Table.~\ref{table:single_qubit_gate} and \ref{table:two_qubit_gate}.
Using these elementary operations on the tableau, 
we employ them to construct a Clifford transformation to diagonalize Pauli terms.
Note that ``diagonal" here means that the tableau only contains Pauli {\it Z} components after the transformation,
which means that the corresponding exponentiation is also diagonal in the computational basis.
A concrete construction, 
which is optimized for classical simulation in the proposed method, will be presented in the next section.

% HXH = Z
\begin{table}
	\caption{
	This table shows the Clifford operations on the binary tableau.
	Each $x_{k,l} and z_{k,l}$ indicates the element in the $k$th row and $l$th column of the {\it X} block and {\it Z} block, respectively. 
	In the table, $\oplus$ denotes exclusive-OR. 
	}
	\begin{tabular}{|c|c|c|} \hline
	 gate & sign & operation \\ \hline
	 H(t)          & $r_k= r_k \oplus x_{k,t} z_{k,t}$ & swap $x_{k,t}$ and $z_{k,t}$ \\ \hline
	 S(t)          & $r_k= r_k \oplus x_{k,t} z_{k,t}$ & $z_{k,t} = z_{k,t} \oplus x_{k,t} $ \\ \hline
	 CNOT(c,t) & $ r_k = r_k \oplus x_{k,c} z_{k,t} (x_{k,t} \oplus z_{k,c} \oplus 1) $ 
							& $x_{k,t}=x_{k,t} \oplus x_{k,c}$ and $z_{k,t} = z_{k,t} \oplus z_{k,c} $ \\ \hline
	 CZ(c,t)     & $ r_k = r_k \oplus x_{k,c} x_{k,t} (z_{k,t} \oplus z_{k,c} ) $ 
						  & $z_{k,t}=z_{k,t} \oplus x_{k,c}$ and $z_{k,c} = z_{k,c} \oplus x_{k,t} $ \\ \hline 
	\end{tabular}
	\label{gate_table}
\end{table}

\section{Simultaneous Diagonalization for Partitioned Hamiltonian} \label{sec:simultaneous_diagonalization}
Here we explain our main idea to accelerate the classical simulation of quantum dynamics by partial diagonalization. 
In the following, we describe our algorithm of partial diagonalization. 

\subsection{Preparation} \label{subsec:prepare_sim_diag}
First, we need to partition the given Hamiltonian terms into mutually commuting terms. 
As we mentioned in Sec.~\ref{subsec:partitioning_hamiltonian}, 
we take a greedy approach to do the partitioning.
Then, we transform the partitioned Hamiltonian into binary representation tableau as in Sec.~\ref{subsec:binary_tableau}.
We take a maximum size of the set of linearly independent binary strings. 
To search the set, 
we zero out the lower-triangular matrix of the tableau by applying exclusive-OR between rows. 
The detail of the algorithm is described in Appendix~\ref{appendix:searching_linear_independent_terms}.
As a result, 
we obtain a set of linearly independent Pauli strings in $P_i$.
Note that the binary strings which are obtained from the procedure described in Appendix~\ref{appendix:searching_linear_independent_terms} are 
not directly used in the following process, since they are not the same as the original mutually commuting Pauli strings $P_i$ via the transformation.
The binary tableau consists of $n$ rows and $(2n+1)$ columns. 
If the rank of the tableau is less than $n$, 
we fill the rest of the rows in the tableau with zeros.
We use this tableau to find a series of Clifford transformation to simultaneously diagonalize mutually commuting terms $P_i$.

\subsection{Partial Diagonalization of Hamiltonian} \label{subsec:simul_diag_alg}
Here we explain our algorithm for finding a series of Clifford transformations 
to simultaneously diagonalize the mutually commuting terms. % as follows.
Since the diagonalized terms are represented by Pauli {\it Z} and {\it I}, 
we transform the binary tableau so that all the elements in the {\it X} block are zero. 
Namely, we will search a series of Clifford transformations to zero out all the elements in the {\it X} block. 
The outline of our algorithm is described in Algorithm~\ref{alg:finding_Clifford}.

\begin{enumerate}[(i)]
\item {\it Maximizing the rank of the X block}

We will maximize the rank of the {\it X} block by applying {\it H} gates. 
The procedure is described in Algorithm \ref{alg:maximizing_rank_xblock}.
Using this algorithm,
we compare the rank of the {\it X} block 
with that applied by {\it H} gate on the $i$th target qubit.
If the latter rank is larger than the former, 
we decide to apply the {\it H} gate on the $i$th target qubit. 
When calculating the rank, 
we use an algorithm similar to the algorithm of searching linearly independent binary strings as in Appendix~\ref{appendix:searching_linear_independent_terms}.
Here we need to consider the {\it X} block only instead of the whole tableau. 
We repeat this operations for $i=1$ to $i=n$.

\item {\it Zeroing out the upper triangular matrix of the X block}

We will zero out the upper triangular matrix of the {\it X} block by CNOT gates. 
The process is described in Algorithm~\ref{alg:zeroing_out_upper_X}.
If the $i$th diagonal element of the {\it X} block is zero, 
we search and swap the row with the $j (>i)$th row whose $i$th column element is one. 
If the search and swap are performed, the element of the $i$th row with the $i$th column is one.
Then, if the element of the $k(>i)$th column in the $i$th row represents one, 
we zero out it with CNOT$(i, k)$. 
We repeat this operations for $i=1$ to $i=n$.

\item {\it Zeroing out the Z block}

We will zero out the {\it Z} block by CZ gates and {\it S} gates. 
More concretely, CZ gates are used to zero out the non-zero elements in the non-diagonal elements of the {\it Z} block and 
{\it S} gates are used to zero out the non-zero elements in the diagonal elements of the {\it Z} block.
We describe the procedure in Algorithm~\ref{alg:zero_out_Zblock}.
If the $j (\neq i)$th column of the $i$th row in the {\it Z} block is one, 
CZ$(i, j)$ is applied to zero out the element. 
Then, if both the $i$th diagonal elements of the {\it X} block and the {\it Z} block are ones, 
we apply {\it S} gate on the $i$th qubit. 
We repeat this process for $i=1$ to $i=n$, 
then the {\it Z} block becomes zero matrix.

\item {\it Turning the Pauli {\it X} components into the Pauli {\it Z} components}

By the previous process, the non-zero elements only exists in the {\it X} block. 
We transform the Pauli {\it X} components into Pauli {\it Z} components by {\it H} gates.
More precisely, if the elements of the $i$th column in the {\it X} block has a non-zero element,
we apply {\it H} gate on the $i$th qubit.
We repeat it for $i=1$ to $i=n$, and then the tableau only has the Pauli {\it Z} components. 
The process is described in Algorithm~\ref{alg:PauliX_to_PauliZ}.
\end{enumerate}
In this way, we can simultaneously diagonalize mutually commuting terms in the partitioned Hamiltonian. 
Note that 
to reduce the number of operations in the above procedure (ii) and (iii), 
we omit the operations if all the terms on $i$th qubit in the group consist of Pauli {\it Z} and {\it I} or Pauli {\it X} and {\it I} components. 
This is because if all the terms on $i$th qubit in the group consist of  Pauli {\it Z} and {\it I}, the terms have already partially diagonalized. 
If all the terms on $i$th qubit in the group consist of Pauli {\it X} and {\it I}, 
those can be partially diagonalized by applying $H_i$.

\begin{figure}
	\centering
		\includegraphics[width=0.8\linewidth]{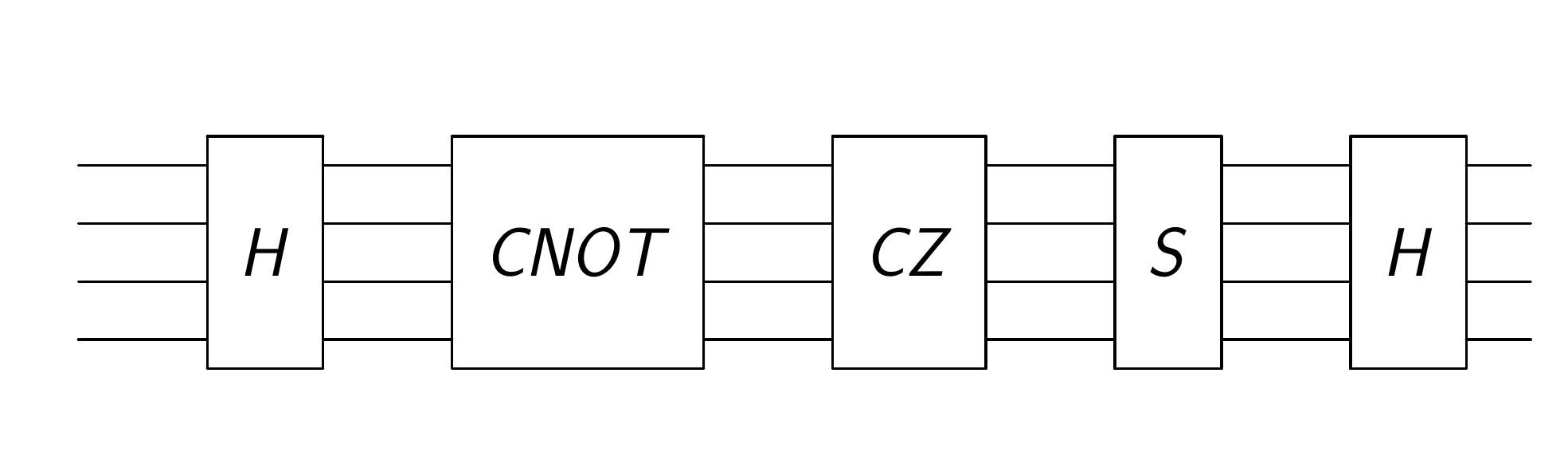}
	\caption{
	This figure shows the quantum circuit made by our algorithm. 
	The quantum circuit consists of a series of {\it H}, CNOT, CZ, {\it S}, and {\it H} gates. 
	In detail, the first {\it H} gates are used in the procedure (i) in our proposed method to maximize the rank of {\it X} block. 
	The CNOT gates are used to zero out the upper triangular matrix of the {\it X} block in the procedure (ii). 
	The CZ and {\it S} gates are used to zero out the {\it Z} block in the procedure (iii). 
	Since these gates do not depend on the order, 
	these gates can be arranged in a series of CZ gates and {\it S} gates.
	The last {\it H} gates are used to transform the Pauli {\it X} to {\it Z} in the procedure (iv).
	}
	\label{fig:clifford_block}
\end{figure}

\begin{algorithm}
\caption{Finding the Clifford transformation of simultaneously diagonalizing mutually commuting terms}
\label{alg:finding_Clifford}
\begin{algorithmic}[1]
\FOR{each group of mutually commuting terms}
\STATE Search the linearly independent binary strings in the group $P_j$.
\STATE Make a binary representation tableau with the linearly independent binary strings. 
\STATE (i) Maximizing the rank of the {\it X} block by {\it H} gates
\STATE (ii) Zeroing out upper triangular matrix of the {\it X} block by CNOT gates
\STATE (iii) Zeroing out the {\it Z} block by CZ and {\it S} gates
\STATE (iv) Turning the Pauli {\it X} components into the Pauli {\it Z} components by {\it H} gates
\ENDFOR
\end{algorithmic}
\end{algorithm}

\begin{algorithm}
\caption{Maximizing the rank of the {\it X} block}
\label{alg:maximizing_rank_xblock}
\begin{algorithmic}[1]
\FORALL {$i$ such that $1 \leq i \leq n$ }
\STATE $r \leftarrow$ the rank of the {\it X} block
\STATE $r^\prime \leftarrow$ the rank of the {\it X} block if we apply ${\it H}(i)$
\IF {$r < r^\prime$}
\STATE we apply ${\it H}(i)$
\ENDIF
\ENDFOR
\end{algorithmic}
\end{algorithm}

\begin{algorithm}
\caption{Zeroing out upper triangular matrix of the {\it X} block}
\label{alg:zeroing_out_upper_X}
\begin{algorithmic}[1]
\FORALL {$i$ such that $1 \leq i \leq n$ }
\IF {the element of $i$th row and $i$th column in the {\it X} block is zero}
\IF {we find the element of $j(>i)$th row and $i$th column in the {\it X} block is one}
\STATE we swap the $i$th row with the $j$th row.
\ENDIF
\ENDIF
\FORALL {$j$ such that $i < j \leq n$ }
\IF{the element of the $i$th row and $j$th column is one}
\STATE we apply CNOT$(i, j)$.
\ENDIF
\ENDFOR
\ENDFOR
\end{algorithmic}
\end{algorithm}

\begin{algorithm}
\caption{Zeroing out the {\it Z} block}
\label{alg:zero_out_Zblock}
\begin{algorithmic}[1]
\FORALL {$i$ such that $1 \leq i \leq n$ }
\FORALL {$j$ such that $1 \leq j \leq n$ }
\IF {the element of the $i$th row and the $j (\neq i ) $th column in {\it Z} block is one}
\STATE we apply CZ$(i, j)$.
\ENDIF
\ENDFOR
\IF {the element of $i$th row and $i$th column in the {\it Z} block is one}
\STATE we apply {\it S}$(i)$.
\ENDIF
\ENDFOR
\end{algorithmic}
\end{algorithm}

\begin{algorithm}
\caption{Turning the Pauli {\it X} components into the Pauli {\it Z} components}
\label{alg:PauliX_to_PauliZ}
\begin{algorithmic}[1]
\FORALL {$i$ such that $1 \leq i \leq n$ }
\IF {the elements of the $i$th column in the {\it X} block has a non-zero element}
\STATE we apply {\it H}$(i)$.
\ENDIF
\ENDFOR
\end{algorithmic}
\end{algorithm}

The Clifford transformation made by our algorithm consists of a series of {\it H}, CNOT, CZ, {\it S}, and {\it H} gates, as shown in Fig.~\ref{fig:clifford_block}. 
In concrete, 
the {\it H} gates act on different target qubits, and the {\it S} gates also.
The CNOT gates are arranged in ascending order of control qubits. 
In addition, the CNOT gates with the same control qubits are arranged in ascending order of target qubits.
The CZ gates are also arranged in the same manner. 
The simulation of applying Clifford transformation can be faster 
by simultaneously applying the CNOT gates with the same control qubits, 
CZ gates, and {\it S} gates, respectively. 
These gates can be implemented in the local operation, 
such as diagonal operators and the operation of swapping the two amplitudes.
We will explain our implementation in Sec.~\ref{subsec:implementation}. 

%%%
\begin{figure}
	\centering
		\includegraphics[width=0.95\linewidth]{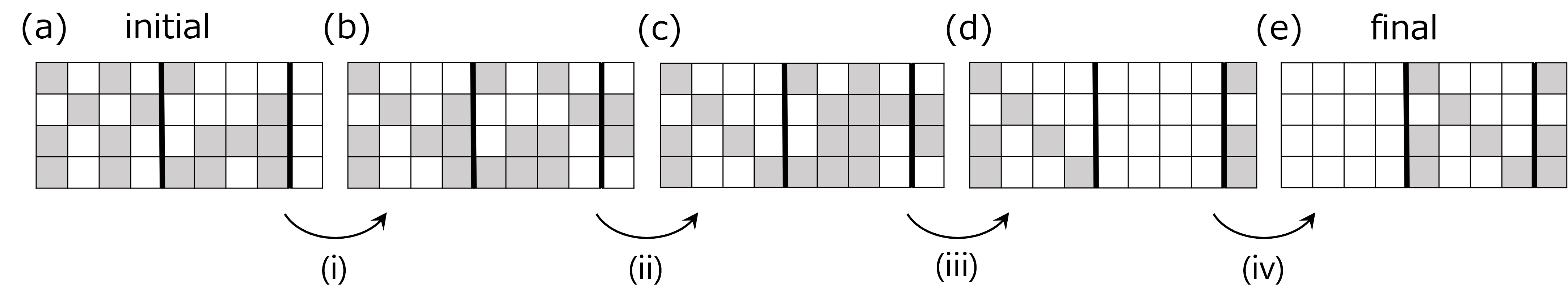}
	\caption{
	This figure shows the process of simultaneous diagonalization. 
	The initial tableau (a) shows the binary representation of $Y_1 I_2 X_3 I_4, I_1 X_2 I_3 Y_4, X_1 Z_2 Y_3 Z_4$ and $Y_1 Z_2 X_3 Z_4$. 
	The binary representation of these terms are $10101000$, $01010001$, $10100111$, and $10101101$, respectively. 
	The white and gray cells represent zero and one in the table, respectively.
	After step (i) of maximizing the rank of {\it X} block with applying Hadamard gate on $3$rd and $4$th qubit, the tableau becomes (b). 
	In step (ii), to zero out the upper triangular matrix of the {\it X} block, CNOT$(2,4)$, and CNOT$(3,4)$ are applied. 
	The tableau goes (c). 
	To zero out the {\it Z} block in step (iii), CZ$(1,3)$, CZ$(2,3)$, CZ$(2,4)$, {\it S}$(1)$, and {\it S}$(2)$ are applied. The tableau becomes (d). 
	Finally, after applying {\it H} gates on all qubits in step (iv), the terms are diagonalized, as in (e). 
	}
	\label{fig:sim_diag}
\end{figure}

Let us consider a concrete example of diagonalizing mutually commuting terms as 
\begin{equation} \label{eqn:eg_diag_hami}
H= Y_1 I_2 X_3 I_4 + I_1 X_2 I_3 Y_4 + X_1 Z_2 Y_3 Z_4 + Y_1 Z_2 X_3 Z_4 
+ X_1 I_2 Y_3 I_4 + I_1 Y_2 I_3 X_4 + Z_1 X_2 Z_3 Y_4 + Z_1 Y_2 Z_3 X_4.
\end{equation}
In this Hamiltonian, one of the sets of linearly independent terms 
is $Y_1 I_2 X_3 I_4 , I_1 X_2 I_3 Y_4 , X_1 Z_2 Y_3 Z_4 $ and $Y_1 Z_2 X_3 Z_4 $.
These binary representation are $10101000$, $01010001$, $10100111$, and $10101101$, respectively. 
We make the tableau of these terms as Fig.~\ref{fig:sim_diag} (a). 
In the figure, we describe zero as a white cell and one as a gray cell. 
The rest of Fig.~\ref{fig:sim_diag} shows how the binary tableau is transformed under the procedures above. 
In concrete, 
in the procedure (i) to maximize the rank of the {\it X} block, 
${\it H}(3)$ and ${\it H}(4)$ are applied. 
Next, CNOT$(2,4)$ and CNOT$(3,4)$ are applied to zero out the upper triangular matrix of the {\it X} block in the procedure (ii). 
Then, CZ$(1,3)$, CZ$(2,3)$ and CZ$(2,4)$ are applied to zero out the non-diagonal element of the {\it Z} block, 
and ${\it S}(1)$ and ${\it S}(2)$ are applied to zero out the diagonal element of the {\it Z} block in the procedure (iii). 
Finally, the Pauli {\it X} components are transformed into Pauli {\it Z} components by ${\it H}(1)$, ${\it H}(2)$, ${\it H}(3)$, and ${\it H}(4)$ in the procedure (iv). 
The transformation $C$ is written by the following equation
\begin{eqnarray*}
C = (H_1 H_2 H_3 H_4) (S_1 S_2)( \text{CZ}(1,3) \text{CZ}(2,3) \text{CZ}(2,4) )
( \text{CNOT}(3,4) \text{CNOT}(2,4))(H_3 H_4) 
\end{eqnarray*}
Thus, the Clifford transformation consists of a series of {\it H}, CNOT, CZ, {\it S}, and {\it H} gates. 
By this transformation, the Hamiltonian written in Eq.~\eqref{eqn:eg_diag_hami} is diagonalized as 
\begin{equation*}
H = C^\dagger (
- Z_1 + Z_2 - Z_1 Z_3 - Z_1 Z_4 
- Z_1 Z_3 Z_4 + Z_2 Z_4 + Z_2 Z_3 Z_4 + Z_2 Z_3 
)C. 
\end{equation*}
We will explain how such diagonalization can accelerate classical simulation in the following sections.

\subsection{Classical Simulation of Quantum Dynamics with Simultaneous Diagonalization} \label{subsec:sim_qd_simul_diag}
% Introduction
In this section, we describe how to perform classical simulation by our proposed method. 
We diagonalize the Hamiltonian terms 
by the Clifford transformation $C_j$,
which has been found in the previous argument, for $j$th group of mutually commuting terms $P_j$ as
\begin{equation*}
P_j = \sum_{k=1}^{m_j} c_{j,k} C_j^\dagger (-1)^{r_{j,k}} \Lambda_{j,k} C_j, 
\qquad \text{where} \; r_{j,k} \in \{0,1\}, \Lambda_{j,k} \in \{ {\it I}, {\it Z} \}^{\otimes n}.
\end{equation*}
In the above equation,
$r_{j,k}$ is the $r_k$ in the binary representation tableau of $j$th group, 
and $\Lambda_{j,k}$ is a diagonal operator of $k$th term in $j$th group. 
By the Clifford transformation, the whole Hamiltonian $H$ is described by 
\begin{equation} \label{eqn:hami_sim_diag}
H = \sum_{j=1}^{n_g} P_j 
= \sum_{j=1}^{n_g} C_j^\dagger \left ( \sum_{k=1}^{m_j} (-1)^{r_{j,k}} c_{j,k} \Lambda_{j,k} \right) C_j
\eqqcolon \sum_{j=1}^{n_g} C_j^\dagger \Lambda_j C_j, 
\end{equation}
where $\Lambda_j$ is a diagonal operator of $j$th group 
and $n_g$ is the number of groups of mutually commuting terms.
Note that 
each diagonalized operator $\Lambda_{j,k}$ is obtained 
by applying the Clifford operation to the binary tableau consisting of all the terms in the group. 

% Main idea(method)
To perform classical simulation, we use Trotter-Suzuki decomposition as in Sec.~\ref{subsec:classical_simulation_qd}.
The quantum state at time $t$ time-evolved 
under the Hamiltonian is described by
\begin{equation} \label{eqn:trotter_simul_diag}
|\psi(t) \rangle = e^{-iHt} |\psi(0) \rangle
 = \lim_{n_{TS} \to \infty} \left( \Pi_{j=1}^{n_g} e^{-i P_j (t/n_{TS})} \right)^{n_{TS}} | \psi(0) \rangle
 \simeq \left( \Pi_{j=1}^{n_g} C_j^\dagger e^{-i \Lambda_j \Delta t} C_j \right)^{n_{TS}} | \psi(0) \rangle .
\end{equation}
The partial diagonalization of the Hamiltonian allows us to update the quantum state by Clifford transformation and the exponentiation of diagonal operators. 
We can accelerate the classical simulation,
if the simulation cost for the Clifford transformation is much smaller than
that reduced by diagonalization of the commuting Pauli terms for minimizing the memory accesses.
As we will see below, implementations of Clifford transformations and diagonal operators 
are further optimized to reduce the computational cost.

\subsection{Implementations}
\label{subsec:implementation}
% 全体に対するイントロ
To perform the classical simulation, 
we use one of the fastest quantum circuit simulators,  ``qulacs''\cite{suzuki2020qulacs}. 
In ``qulacs,''
each single gate of {\it H}, {\it S}, CNOT, and CZ gates is implemented, 
but these implementations are not optimal for our method.
In our method, the same kind of gates appears continuously, 
such as consecutive {\it S} gates, CZ gates, CNOT gates with the same control qubit, 
and the exponentiation of diagonal operators in $\Lambda_j$ in Eq.~\eqref{eqn:hami_sim_diag}.
So, we implement them with the same concept of implementation in ``qulacs.''

\begin{enumerate}
\item {\it CNOT gates}

% introduction
Here we explain how to update a state vector by the consecutive CNOT gates with the same control qubit. 
Since the update of the state vector by these operations is similar to the update by a single CNOT gate, 
we will first describe the update of the quantum state by a single CNOT gate. 
% implementation of single CNOT gate
The matrix of CNOT gate is written in Eq.~\eqref{eqn:clifford_gates}.
The implementation is realized by swapping the amplitudes between the $|10\rangle$ and $|11\rangle$.
The control qubits of both basis states are one 
and the relation between the target qubits of the basis states is bit-flipped.
This is easily expanded to the multiple CNOT gates with the same control qubit.
% implementation of multiple CNOT gates
Let us consider the simulation of applying CNOT$(c,t_1)$, CNOT$(c, t_2)$, $\ldots$, and CNOT$(c, t_k)$ gates to the quantum state.
The implementation is to swap the amplitudes between basis states 
whose $c$th qubits are both one and bitstrings consisting of $t_1, t_2, \ldots, t_k$ th qubits have a relation of bit-flip. 
More concretely, 
let us denote the bitstring of basis state as $b_n \ldots b_1$, 
then those bitstrings consisting of target qubits are written by $b_{t_1} b_{t_2} \ldots b_{t_k}$ and $(b_{t_1} \oplus 1) (b_{t_2} \oplus 1) \ldots (b_{t_k} \oplus 1)$.

\item {\it S and CZ gates}

% introduction
Here we explain the implementation of simultaneously applying {\it S} and CZ gates. 
% multiple S gates
First, we describe an implementation of applying multiple {\it S} gates.
As we described in Eq.~\eqref{eqn:clifford_gates},
{\it S} gate is a diagonal matrix.
Let us consider the simulation of applying ${\it S}(t_1)$ and ${\it S}(t_2)$.
We take a bitstring consisting of the target qubits $t_1$ and $t_2$ of the basis states.
By counting the ones in the bitstring,
we can calculate the coefficient.
For each basis state $a_{b_n \ldots b_1} |b_n \ldots b_1 \rangle,$ 
where $a_{b_n \ldots b_1} \in \mathbb{C}, b_j \in \{ 0, 1 \}$, 
the amplitudes $a_{b_n \ldots b_1}$ are updated by multiplying $i^{b_{t_1}+b_{t_2}}$.
Similarly, we can simultaneously apply more than two {\it S} gates to the quantum state. 
Let us denote the target qubits $t_k$, 
we update the amplitudes $a_{b_n \ldots b_1}$ by multiplying $i^{\sum_k b_{t_k}}$.
%  multiple CZ gates
Next, we explain an implementation of applying multiple CZ gates.
As we described in Eq.~\eqref{eqn:clifford_gates},
the CZ gate is also a diagonal matrix.
The implementation of the single CZ gate on control qubit $c$ and target qubit $t$ is to multiply $-1$ to the amplitudes 
whose basis states are both ones on the control qubit $c$ and target qubit $t$. 
When applying multiple CZ gates, for each basis states, 
we compute the coefficient over all CZ gates by the same operation as a single CZ gate in advance.
In concrete, 
by denoting the control and target qubits of $j$th CZ gate as $c_j$ and $t_j$,
and the bits of the control and target qubits of the basis state as $b_{c_j}$ and $b_{t_j}$, 
the coefficient is written by $(-1)^{\sum_j b_{c_j} b_{t_j} }$.
Then we update each amplitudes by multiplying it.
In this way, we can reduce the number of memory access to the state vector.
% multiple S and CZ gates
Since the implementation of {\it S} and CZ gates is based on counting the number of ones in the bitstring consisting of target qubits and control qubits of basis states, 
we can simultaneously apply all {\it S} and CZ gates to each basis state 
by calculating the coefficients and then applying them to the amplitudes. 

\item {\it diagonal operator $\Lambda_j$}

Here we explain how to efficiently update the quantum state 
by the diagonal operator $\Lambda_j$. 
By the partial diagonalization, the terms in the partitioned Hamiltonian are written by the $n$-fold tensor products of Pauli {\it Z} and {\it I}.  
The simulation of a single diagonal operator $(-1)^{r_1} c_1  \otimes_{j=1}^n {\it Z}^{z_j},$ where $r_1, z_j \in \{0,1\}, c_1 \in \mathbb{R}$ is performed as follows.
For each basis state $a_{b_n \ldots b_1} |b_n \ldots b_1 \rangle,$ 
where $a_{b_n \ldots b_1} \in \mathbb{C}, b_j \in \{ 0, 1 \}$, 
we count the ones in the bit string of basis state 
whose indexes are the same as the target qubit indexes of Pauli {\it Z}s in the operator, 
such as $\sum_{j=1}^n b_j z_j$. 
We multiply the amplitudes $a_{b_n \ldots b_1}$ by the coefficient $e^{-i \Delta t c_1 (-1)^{r_1+\sum_{j=1}^n b_j z_j}}$. 
Similarly, 
the simulation of exponentiation of multiple diagonal operators is to calculate the coefficients for all the terms in the partitioned Hamiltonian, 
and then read the amplitudes and multiply the amplitudes by the coefficients.
In concrete, let us consider the diagonal operator $\Lambda_j$ in Eq.~\eqref{eqn:hami_sim_diag}, such as
\[
\Lambda_j = \sum_{k=1}^{m_j} (-1)^{r_{j,k}} c_{j,k} \left ( \otimes_{l=1}^n {\it Z}^{z_{k,l}} \right ), \quad \text{where} \; z_{k,l} \in \{ 0, 1 \}.
\]
We multiply the amplitudes by 
\[
e^{-i \Delta t \sum_{k=1}^{m_j} c_{j,k} (-1)^{ r_{j,k} + \sum_{l=1}^n z_{k,l} b_l} }
\]
for each basis state $a_{b_n \ldots b_1} |b_n \ldots b_1 \rangle,$ 
where $b_l \in \{ 0, 1 \},$ for $l=1,\ldots,n$ and $a_{b_n \ldots b_1} \in \mathbb{C}.$
\end{enumerate}

Using these implementations, 
we reduce the number of memory access to the state vector.
The number of memory access  
is $O(n 2^n)$ by {\it H} gates, $O(2^n)$ by {\it S} and CZ gates, $O(n 2^n)$ by CNOT gates, 
and $O(2^n)$ by an exponentiation of diagonal operator $\Lambda_j$. 
The total cost is $O(n 2^n)$ for each group,
and then $O(n_g n 2^n)$ for all groups. 
On the other hand, the simulation cost 
for applying each Pauli term in the Hamiltonian independently
is $O(m 2^n)$ with $m$ being the number of Pauli terms.
Therefore, our proposed method works well 
if the number of groups $n_g$ is $1/n$ times smaller than that of the terms $m$. 
For example, in the case of a fully connected transversal Ising model,
$n_g=2$ and $m=O(n^2)$ and hence we can expect acceleration by the proposed method.
For more complicated Hamiltonian with fermions, 
whether or not we can obtain any acceleration highly depends on the number of groups and cache efficiency, 
since the effect of the acceleration decays with the number of groups 
and the limited size of cache causes delay by the memory access to obtain coefficients or amplitudes. 

Note that it is easy to parallelize the calculation of multiplying the state vector by diagonal matrices as {\it S} gate, CZ gate and exponentiation of diagonal operator $\Lambda_j$
because the calculation can be performed independently on each basis state. 
Similarly, the calculation of multiple CNOT gates with same control qubit depends only on pairs of two basis states, 
so it is also easy to perform the simulation in parallel.
By the parallelization, 
the simulation is performed in $O(n_g n \lceil 2^n/n_{\text{thread}} \rceil)$, 
where $n_{\text{thread}}$ denotes the number of threads and $\lceil 2^n/n_{\text{thread}} \rceil$ represents the number of cycles of the parallel execution.
Similarly, the simulation for applying each Pauli term in the Hamiltonian is performed in $O(m \lceil 2^n/n_{\text{thread}} \rceil)$.
So, the simulation time will be reduced by a factor of $O(n n_g/m)$.

\section{Numerical Experiments}\label{sec:numerical_experiments}
In this section, the numerical experiments are performed 
to validate the acceleration of classical simulation by our proposed method.
We perform the classical simulation of Hamiltonian dynamics under the Trotter-Suzuki decomposition with the time step $\Delta t =t/m= 0.01$ according to Eq.~\eqref{eqn:trotter_decomp} and Eq.~\eqref{eqn:trotter_simul_diag}. 
We compare the simulation time of one Trotter step with and without our proposed method for two models: 
the transverse field Ising model and SYK Majorana fermion model \cite{sachdev1993gapless,kitaev2015simple}. 
The former Hamiltonian is used in quantum algorithms such as quantum annealing \cite{kadowaki1998quantum} and quantum approximated optimization algorithm \cite{farhi2014quantum}.
The latter one is a prototypical model for condensed matter physics describing a strongly correlated electron system. 
It shows interesting phenomena such as quantum chaos and quantum information scrambling \cite{iyoda2018scrambling, maldacena2016remarks}.

To perform the numerical experiments, 
we partition the given Hamiltonian into groups of mutually commuting terms, 
and find the Clifford transformations to diagonalize each group.
The former Hamiltonian is easily partitioned with the minimum number of groups. 
Also, the Clifford transformation of simultaneously diagonalizing each group is easily found. 
On the other hand, the partition of the latter Hamiltonian with the minimum number of groups 
and Clifford transformation for them 
are not easily found by hand, 
so we have to systematically find them with the algorithm as we described in Sec.~\ref{subsec:simul_diag_alg}.
By the partitioned Hamiltonian and the Clifford transformation, 
we update the quantum state by simultaneously applying terms in each group to accelerate the simulation. 

It is important whether the coefficients in the groups can be stored in the cache. 
In the case of the transverse field Ising model, the number of terms in the Hamiltonian is $465$ at $30$ qubits. 
Since our CPU has $32$kB of the L1 cache, 
which corresponds to four thousand floating points with double precision,
the L1 cache can store all coefficients. 
Therefore, the simulation is expected to be performed fast. 
In the case of the Majorana fermion models, 
the number of terms at 28 qubits is about 367,000. 
These terms with double precision floating points occupy 3MB, 
which is more than the size of the L2 cache (1MB) but less than that of the L3 cache (11MB). 
So, the simulation is also expected to be performed fast. 
The effect of the acceleration is less than in the former case 
because the access to the L1 cache is faster than the L3 cache. 
For GPU, the L2 cache has about 6MB, 
so the coefficients of the models can be stored in the L2 cache. 
Since GPU has large bandwidth and thousands of cores, it is expected to perform simulation faster than CPU. 
Note that the amplitudes of the exponentially large state vector are also stored in the cache during the simulation. 
If the amplitudes are not stored in the cache, 
the data must be obtained through the memory access, 
and then it causes an increase in simulation time.

To confirm the effect of the acceleration, 
we compare the simulation time between the two methods for each model. 
One of them is to apply exponentiation of each Hamiltonian term one by one as Eq.~\eqref{eqn:trotter_decomp}.
The other one is to simultaneously apply exponentiation of terms in each group as Eq.~\eqref{eqn:trotter_simul_diag}.
We call the former method a baseline method and the latter method a proposed method. 
In these simulations, we use the quantum circuit simulator ``qulacs''~\cite{suzuki2020qulacs} with our additional implementation described in Sec.~\ref{subsec:implementation}, 
such as fast Clifford transformations and diagonal operations. 
Note that the implementation for CPU is parallelized by SIMD (Single Instruction Multiple Data) and OpenMP. 
The results are described in the following section.

\subsection{Transverse Field Ising Model} \label{subsec:numexp_ising}
\begin{figure}
	\centering
		\includegraphics[width=0.45\linewidth]{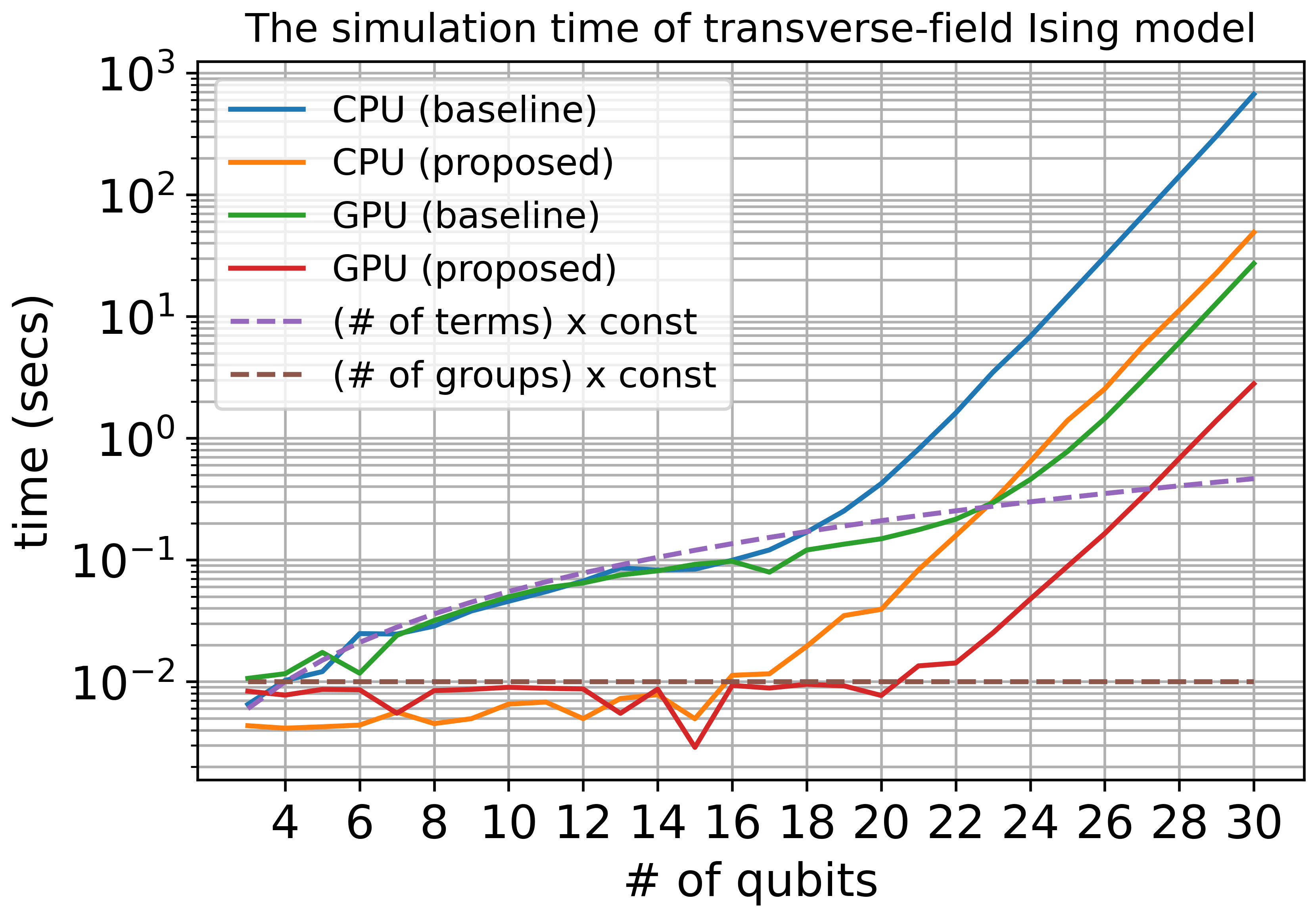}
		\includegraphics[width=0.45\linewidth]{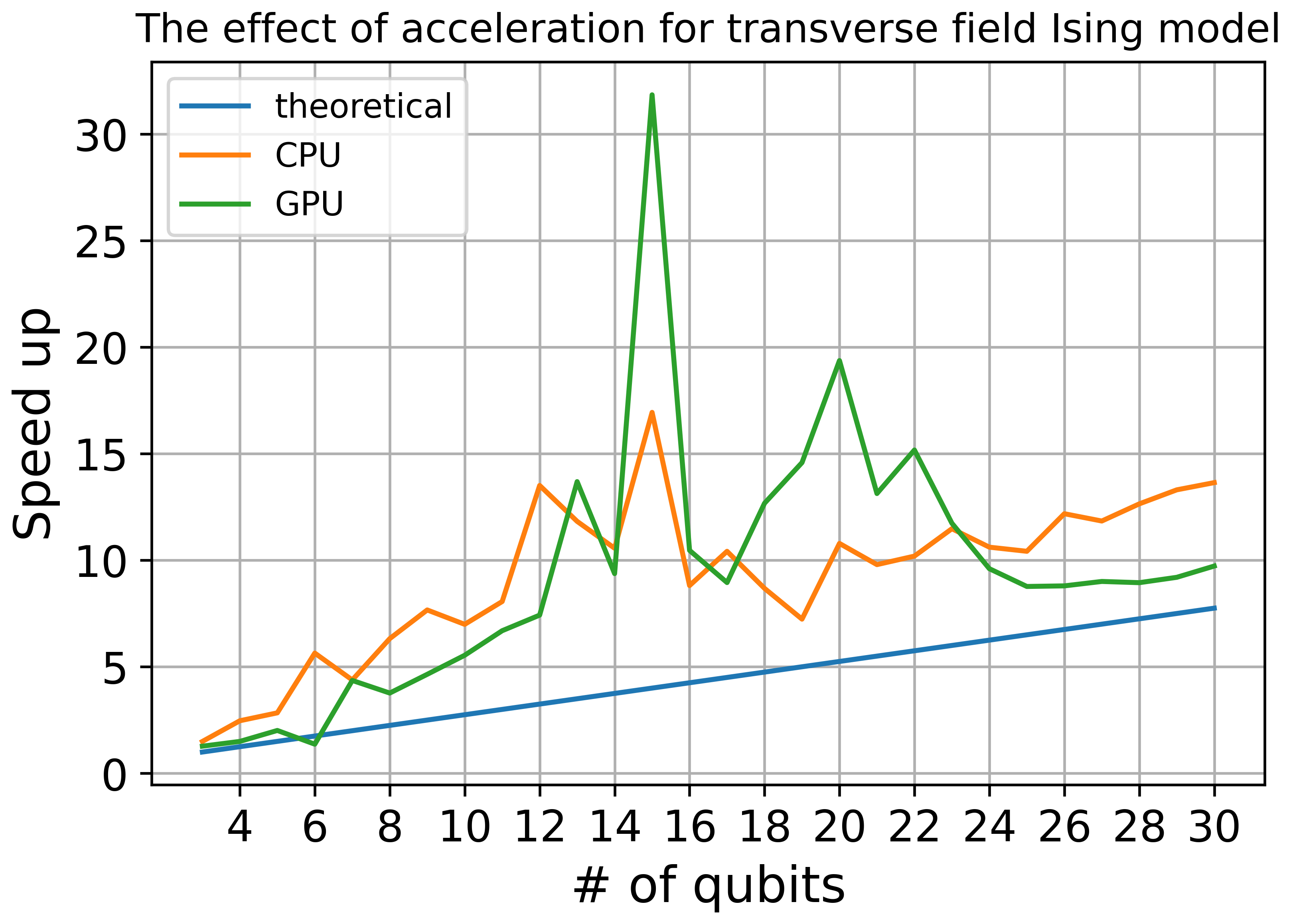}
	\caption{
	The left figure shows the simulation time of transverse-field Ising model on CPU and GPU.
	The right figure shows the speed up between the baseline and the proposed method. 
	The theoretical line represents $m/(n n_g)$. 
	}
	\label{fig:ising_cpu_gpu}
\end{figure}

The first model we will simulate is the transverse field Ising model, 
whose Hamiltonian is given by 
\begin{equation*}
H = \sum_{i < j} J_{ij} Z_i Z_j + \sum_i h_i X_i, 
\end{equation*}
where the $J_{ij}$ and $h_i$ are randomly chosen from the uniform distribution between $-1$ and $1$.
The Hamiltonian term $J_{ij} Z_i Z_j$ is already diagonalized, 
and the term $h_i X_i$ can be diagonalized by $H_i$. 
As we mentioned above, we perform the simulation with two methods: 
multiplying the exponentiation of each Hamiltonian term one by one 
and multiplying the exponentiation of each group of partitioned Hamiltonians. 
The simulation time with both methods on CPU and GPU is shown in Fig.~\ref{fig:ising_cpu_gpu}. 

% CPU 
First, we describe the simulation time on CPU. 
From Fig.~\ref{fig:ising_cpu_gpu}, up to $16$ qubits,
the simulation time of the baseline method is proportional to the number of terms, 
and that of the proposed method is proportional to the number of groups. 
The increase of simulation time over $16$ qubits is more than that below $16$ qubits. 
Since the size of the state vector with $16$ qubits is the same as that of the L2 cache, 
the increase in simulation time is caused by the memory access to the L3 cache or main memory to read the amplitudes of state vector when updating the quantum state. 
The increase of the simulation time over $20$ qubits is much more than that from $16$ to $20$ qubits. 
Since the L3 cache can not include all amplitudes of the state vector over 20 qubits, 
the exponential increase of the simulation time may be caused 
by both the calculation of updating the state vector 
and the memory access to main memory. 

Next, we describe the simulation time on GPU. 
From Fig.~\ref{fig:ising_cpu_gpu}, below $22$ qubits,
the simulation time of the baseline method is proportional to the number of terms 
and that of the proposed method is proportional to the number of groups. 
Since the size of the L2 cache is less than the size of the state vector with $19$ qubits, 
it is expected to delay the simulation over $19$ qubits. 
However, the simulation time below $22$ qubits does not appear to be affected much by the cache size. 
Above $22$ qubits, the simulation time eventually increases exponentially with the number of qubits 
because of both the memory access and the calculation of exponential large state vector.

In terms of the computational cost, 
from $m=O(n^2), n_g=2$, the effect of acceleration is expected by $O(m/(n n_g)) = O(n)$.
The right figure of Fig.~\ref{fig:ising_cpu_gpu} shows the effect of acceleration 
between the baseline and proposed method on CPU and GPU. 
Actually, 
while the cache is effective, the acceleration is much better than expected. 
For example, at 15 qubits, 
since the amplitudes and coefficients can be stored in the L2 cache, 
the simulation is accelerated 30 times faster on GPU and 15 times faster on CPU by our proposed method. 
At $30$ qubits, we have succeeded in performing the simulation $10$ times faster on both CPU and GPU. 

% CPU
% L1 cache (32kB) = 11 qubit < L2 (1024kB) = 16 qubit< 19 qubit (8MB) < L3 cache (11MB) < 20 qubit (16MB)
% GPU
% 18qubit (4096 KB) < L2 cache: 6144 KB < 19qubit (8192 KB)

\subsection{Majorana Fermion Model}
\begin{figure}
	\centering
		\includegraphics[width=0.45\linewidth]{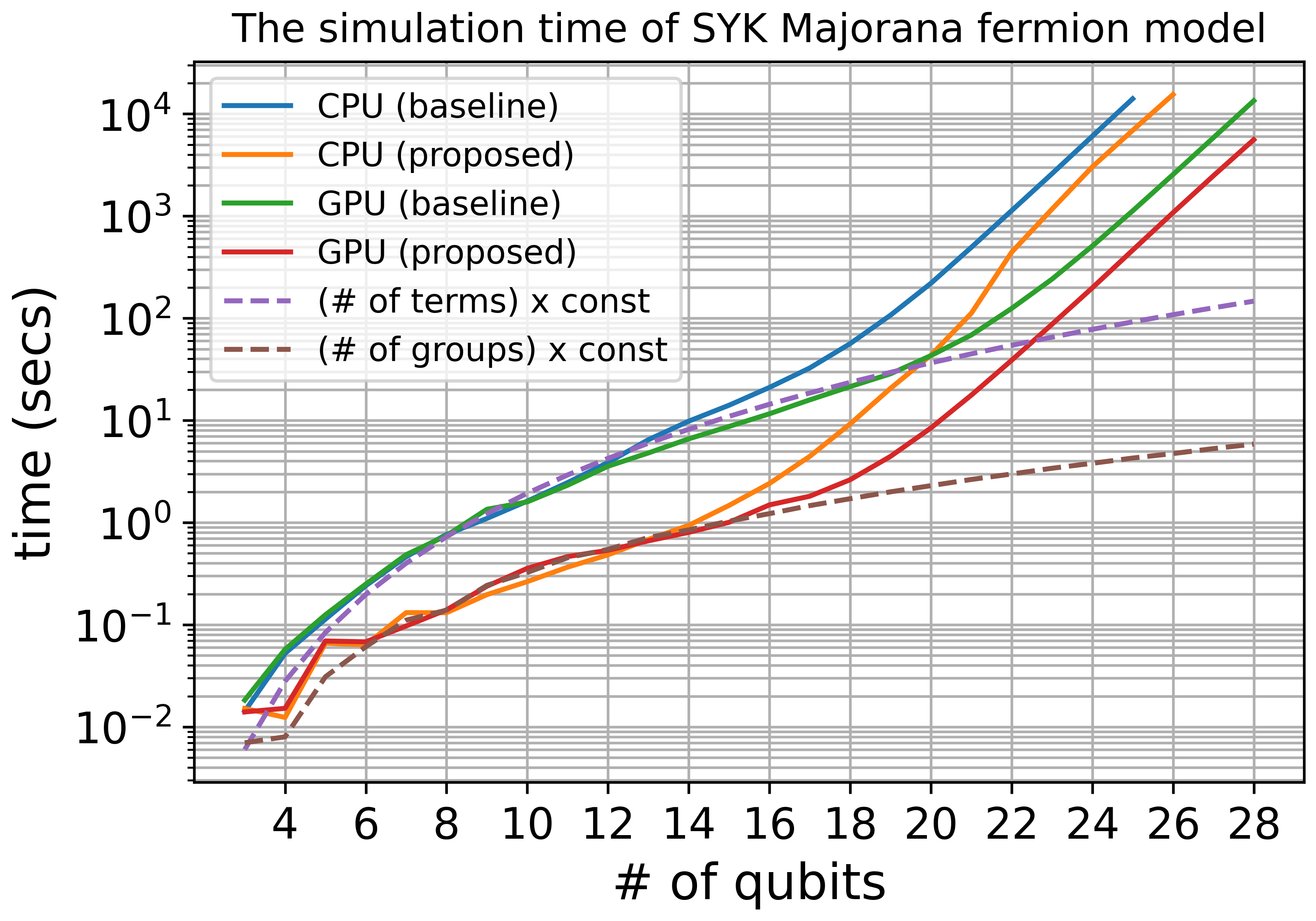}
		\includegraphics[width=0.45\linewidth]{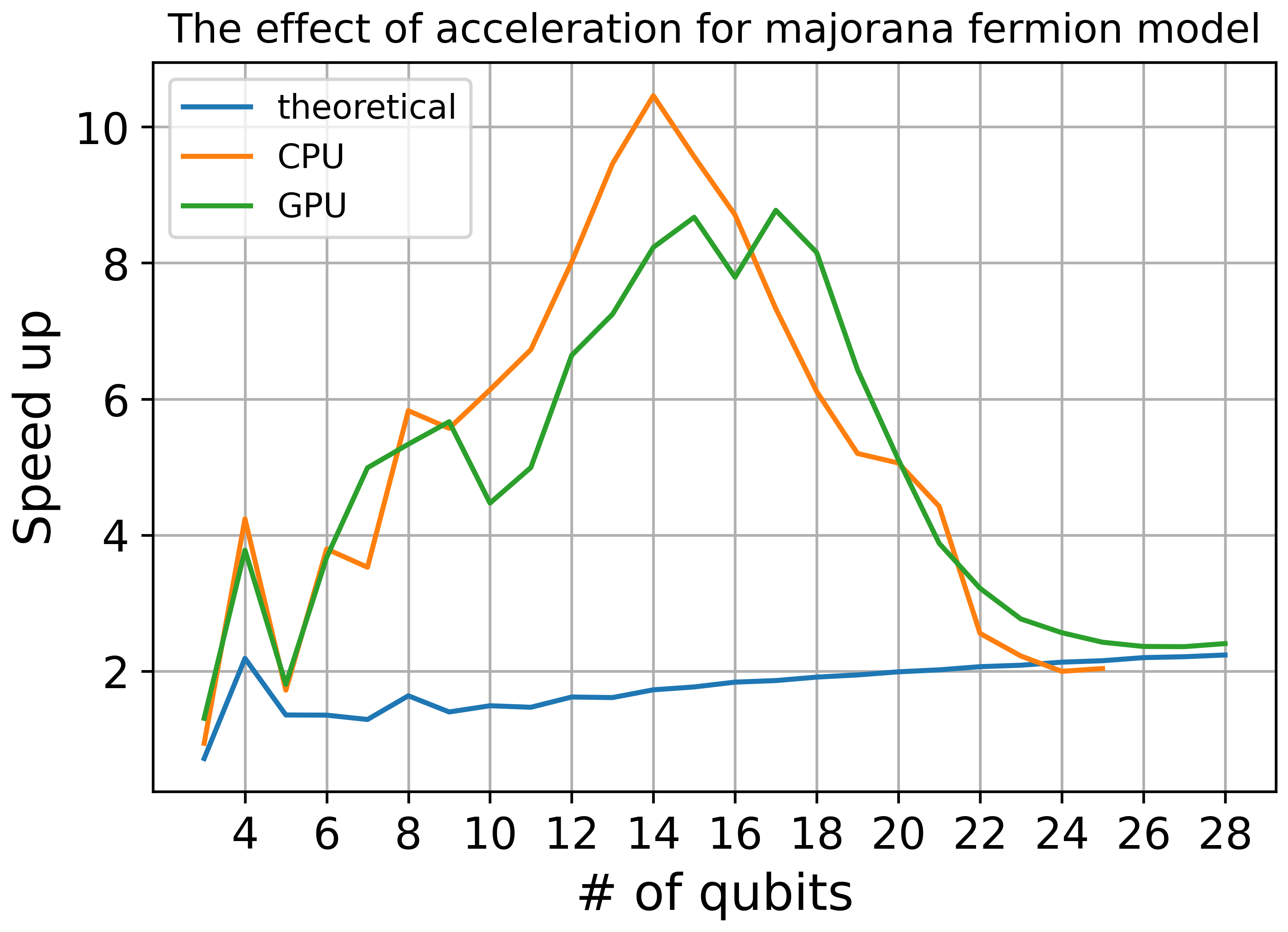}
	\caption{
	The left figure shows the simulation time of SYK Majorana fermion model on CPU and GPU.
	The right figure shows the speed up between the baseline and the proposed method. 
	The theoretical line represents $m/(n n_g)$. 
	}
	\label{fig:majorana_cpu_gpu}
\end{figure}

The second model we will simulate is the Majorana fermion model. 
The Hamiltonian of the SYK (Sachdev-Ye-Kitaev) Majorana fermion model is given by 
\begin{align*}
H &= \sum_{ijkl} J_{ijkl} \hat{c}_i \hat{c}_j \hat{c}_k \hat{c}_l, \\
\hat{c}_{2i-1} &= Z_1 \ldots Z_{i-1} X_i, \, \hat{c}_{2i} = Z_1 \ldots Z_{i-1} Y_i, \\
\{ \hat{c}_i, \hat{c}_j\} &= 2\delta_{ij},
\end{align*}
where $J_{ijkl}$ follows a complex Gaussian distribution, and  $J_{ijkl}=-J_{jikl}=-J_{ijlk}=J_{lkji}^*$ holds. 
As we mentioned above, the groups of mutually commuting terms in the Hamiltonian is not easily found. 
To find them, we use our algorithm in Sec.~\ref{subsec:simul_diag_alg}. 
Then, we measure the simulation time for the two methods on CPU and GPU. 
The results of the simulation is shown in the left figure of Fig.~\ref{fig:majorana_cpu_gpu}. 

First, we describe the simulation time on CPU. 
Similar to the results of the transverse-field Ising model in Sec.~\ref{subsec:numexp_ising}, 
below $16$ qubits, the simulation time of the baseline method is proportional to the number of terms and that of the proposed method is proportional to the number of groups. 
Above $16$ qubits, the increase in simulation time is larger than that below $16$ qubits.  
Since the size of the L2 cache is the same as that of the state vector with $16$ qubits, 
the access to the L3 cache or main memory to read the amplitudes of state vector cause the increase in the simulation time.
Moreover, the size of the state vector at $20$ qubits is larger than that of the L3 cache. 
Actually, the simulation time gradually shift 
from an increase proportional to the number of terms or groups to an exponential increase with the number of qubits.
Above $22$ qubits, the simulation time for both the baseline method and the proposed method exponentially increases with the number of qubits.

Next, we describe the simulation time on GPU.  
The simulation time of the baseline method is proportional to the number of terms 
and that of the proposed method is proportional to the number of groups.
Since the L2 cache can not store whole state vector above $19$ qubits and need to access the global memory to read the amplitudes, 
the increase in simulation time on GPU above $19$ qubits is larger than that below $19$ qubits. 
Above $19$ qubits, the increase in simulation time gradually transitions 
from an increase proportional to the number of terms or groups 
to an exponential increase with the number of qubits. 
Above $22$ qubits, the simulation time for both the baseline method and the proposed method exponentially increases with the number of qubits.

In terms of the computational cost, 
from the right figure in Fig.~\ref{fig:majorana_cpu_gpu},
the expected acceleration effect is about twice faster by our algorithm
because of $m/(n n_g) \simeq 2$. 
Around the $15$ qubits, the effect of the acceleration is about eight times faster by our proposed method. 
Then, the acceleration effect gradually decreases with the number of qubits, 
and, over $22$ qubits, eventually is almost the same as the expectation by $m/(n n_g) \simeq 2$. 
Since the size of the state vector at $16$ qubits is the same as L2 cache and 
that at $20$ qubits is larger than the size of L3 cache,
the acceleration depends on cache efficiency. 
At $25$ qubits, we confirm that our proposed method performs the simulation twice faster than the baseline method for both CPU and GPU.
The effect of acceleration is smaller than the result of the transverse-field Ising model. 
It would be caused by how much the number of groups can be reduced.

% CPU
% L1 cache (32kB) = 11 qubit < L2 (1024kB) = 16 qubit< 19 qubit (8MB) < L3 cache (11MB) < 20 qubit (16MB)
% GPU
% 18qubit (4096 KB) < L2 cache: 6144 KB < 19qubit (8192 KB)

\section{Conclusions} \label{sec:conclusion}
In this paper, we propose a method of accelerating the classical simulation of Hamiltonian dynamics.
Specifically, we partition the Hamiltonian terms into groups of mutually commuting terms, 
and then simultaneously diagonalize the terms in the groups by Clifford transformation.
To verify the effect of acceleration by simultaneous diagonalization, we perform numerical experiments for two models: the transverse-field Ising model and the Majorana fermion model. 
In our numerical experiments, we confirmed that the simulation is performed faster by simultaneous diagonalization than by time-evolving Hamiltonian terms one by one.
We also confirmed the acceleration on both CPU and GPU.
In other words, we confirmed the acceleration is independent of the architecture.

Here we explore possbile applications of our proposed method. 
The simulation of Quantum annealing \cite{kadowaki1998quantum} and Quantum Approximate Optimization Algorithm (QAOA) \cite{farhi2014quantum}
can be accelerated by our algorithm. 
Both algorithms are based on Hamiltonian dynamics of Ising models with transversal fields or other driving Hamiltonians and hence can be accelerated easily.
On the other hand, the fermionic Hamiltonians are often emploied for quantum simulation or valiational quantum eigensolver to solve quantum many-body systems like condensed matter physics and quantum chemistry.
As the number of qubits in quantum computers continues to grow, fast classical simulations are becoming increasingly important to know if these algorithms work properly and/or what parameters to choose to improve their performance. Our proposed method will make a significant contribution to accelerating the development of such quantum algorithms.

%% Loading bibliography style file
\bibliographystyle{model1-num-names}
% \bibliographystyle{model2-names}
% \bibliographystyle{cas-model2-names}

% Loading bibliography database
\bibliography{cas-refs}

\appendix
\section{The algorithm of searching linear independent terms}
\label{appendix:searching_linear_independent_terms}
Here we explain the algorithm for finding a set of linear independent terms from the mutually commuting terms.
The algorithm is described in Algorithm.~\ref{alg:linearly_independent}. 

\begin{figure}
	\centering
		\includegraphics[width=0.75\linewidth]{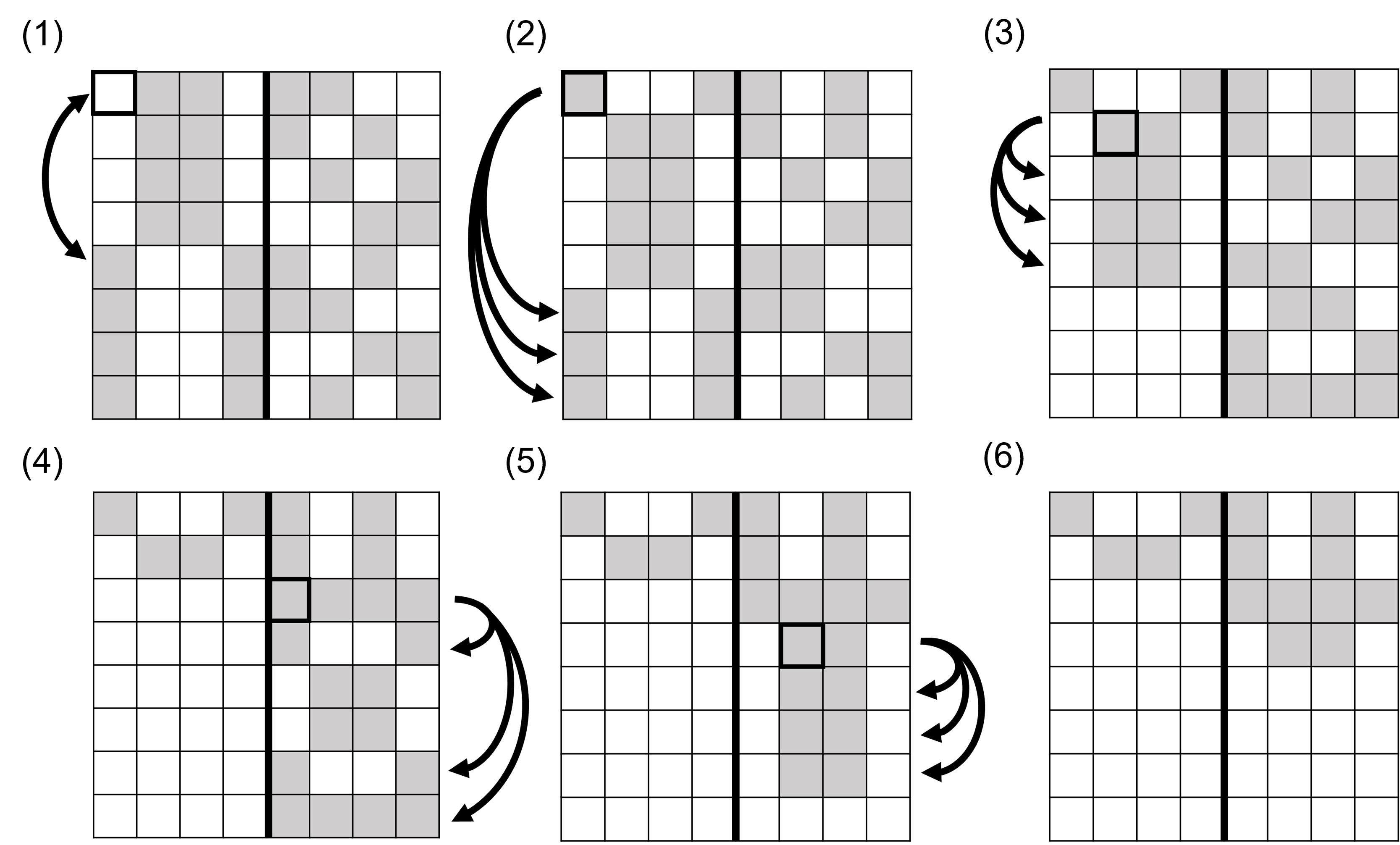}
	\caption{This figure shows an example of the process of searching the linear independent vectors in the binary tableau consisting of mutually commuting terms.}
	\label{fig:searching_linear_independent_vec}
\end{figure}

Let us consider a example of searching linear independent terms of 
$H=ZYXI+ZXYI+IYXZ+IXYZ+YIZX+YZIX+XIZY+XZIY$.
First, we transform the terms into binary strings and make a tableau as we described in Sec.~\ref{subsec:binary_tableau}.
Let us denote the $i_r$ as the row index and
$i_c$ as the column index in the tableau.
We start from the 1st column and 1st row of the tableau.
If the $i_c$ th column element in $i_r$ th row is zero and 
we find the $j(>i_r)$th row whose $i_c$ th column element is one,
we swap the $j$th and $i_r$th rows.
Otherwise, we increase the $i_c$ to $i_{c+1}$ and go back to the process of finding the row whose $i_c$th column element is one.
In Fig.~\ref{fig:searching_linear_independent_vec} (1), we swap the $1$st and $5$th row, because the $1$st column of the $1$st row is zero and that of the $5$th row is one.
Then, we search for the $k(>i_r)$th rows whose $i_c$ th column elements have one.
If we find them, we take exclusive-OR between $i_r$ th row and the $k$th rows.
In Fig.~\ref{fig:searching_linear_independent_vec} (2), 
since the $1$st column element of the $6, 7$ and $8$th rows is one, 
we apply exclusive-OR between the $1$st row and $6, 7$ and $8$th rows.
We increment $i_r$ and $i_c$ to $i_r+1$ and $i_c+1$, respectively.
We repeat it until $i_r=N$ or $i_c=2n$, as in Fig.~\ref{fig:searching_linear_independent_vec} (3), (4), and (5).
The terms corresponding to the rest of the rows are the linear independent terms.
In Fig.~\ref{fig:searching_linear_independent_vec} (6), 
since the rest of terms corresponds to the initial $5$th, $2$nd, $3$rd, and $4$th rows,
the linearly independent terms are $YIZX$, $ZXYI$, $IYXZ$, and $IXYZ$.

\begin{algorithm} 
\caption{Implementation of searching linear independent vectors}
\label{alg:linearly_independent}
\begin{algorithmic}[1]
\REQUIRE A tableau of binary representation of mutually commuting terms
\STATE Let us denote $N$ as the number of terms and $n$ as the number of qubits.
\STATE Let us denote $i_r$ as $i_r$ th rows (terms) in the tableau and $i_c$ as the $i_c$ th columns in the tableau. 
\STATE $i_r \gets 1, i_c \gets 1$.
\WHILE {$i_r \leq N$ and $i_c \leq 2n$}
\IF {the element of $i_c$th column in the $i_r$th row is $0$}
\STATE We search for the $j(>i_r)$th row whose element of $i_c$th column is $1$.
\STATE If we find such a row, we swap the $i_r$th row with the $j$th row. 
\STATE Otherwise we increment $i_c$ as $i_c \gets i_c+1$ and go back to line $4$.
\ENDIF
\IF {the element of $i_c$th column in the $i_r$th row is $1$}
\STATE We delete the ones in the $i_c$th column of $k(< i_r)$th row by exclusive-OR between the $i_r$th row and the $k$th row.
\STATE $i_r \gets i_r+1$, $i_c \gets i_c+1$.
\ENDIF
\ENDWHILE
\RETURN the corresponding terms whose elements in the final binary tableau have at least one non-zero element.
\end{algorithmic}
\end{algorithm}

\end{document}